\documentclass{article}

\begin{document}

% Redefine "plain" pagestyle
%\makeatletter     % `@' is now a normal "letter' for LaTeX
%\renewcommand{\ps@plain}{%
     %\renewcommand{\@oddhead}{\textrm{Your Header}\hfil\textrm{\thepage}}%
     %\renewcommand{\@evenhead}{\@oddhead}%
     %\renewcommand{\@oddfoot}{}% empty footer
     %\renewcommand{\@evenfoot}{\@oddfoot}}
%\makeatother     % `@' is restored as a "non-letter" character

\title{Analytic Proof of the Attractors of a Class of Cellular
Automaton}         % Enter your title between curly braces
\author{Ru-Fen Liu and Chia-Chu Chen}   % Enter your name between curly braces
\date{National Cheng-Kung university,\\ Physics department,\\Tainan, Taiwan}               % Enter your date or \today between curly braces
\maketitle
% Set to use the "plain" pagestyle
\pagestyle{plain}
\begin{abstract}
\indent

In this work we provide analytic results of infinite
one-dimensional cellular automaton($\bf CA$). By realizing
symbolic products, we investigate a subclass of infinite $\bf CA$
and prove analytically that within this subclass the only allowed
attractors are homogenous, steady and periodic states for
arbitrary initial configuration. Our method also provide exact
enumeration of these attractors and it is shown explicitly in a
particular model.

\end{abstract}

\section{Introduction}       % Enter section title between curly braces
\indent \boldmath \ \ \ \ \ \ $\bf Cellular$ $\bf automata(CA)$
are simple mathematical models which can generate complex
dynamical phenomena. In principle, $\bf CA$ are discrete dynamical
systems defined on a discrete lattice. The state of each site at
any time $t$ is in one of the $g$ states. The interaction between
sites is according to given local rules and the system evolves
synchronously in discrete time steps. Cellular automaton was
introduced by von Neumann in order to address the problems of
self-reproduction and evolution \cite{Neu}. However $\bf CA$ has
not attracted much attention until John Conway introduced the game
of lifes \cite{Con} around 1970 by using $\bf CA$. About twenty
years ago, Stephen Wolfram introduced cellular automaton to the
physics community as models of complex dynamical systems
\cite{Wol} and a new approach to the parallel computing scheme(For
a review see \cite{Mit}). The interest in $\bf CA$'s potential
application continues growing. In fact cellular automata have been
used to simulate, for example, solar flares \cite{Isl}, fluid
dynamics \cite{Fri}, crystal growth \cite{Mac}, traffic flow
\cite{Fuk} and galaxy formation \cite{Ger}.
\newline
\indent
         Based on large number of numerical studies, Wolfram
has suggested that $\bf CA$ can be classified into four classes.
Cellular automata within each class has the same qualitative
behavior. Starting from almost all initial conditions,
trajectories of $\bf CA$ become concentrated onto attractors, the
four classes can then be characterized by their attractors.
According to Wolfram's classification, the class 1, 2 and 3 are
roughly corresponding to the limit points, limit cycles and
chaotic attractors in continuous dynamical systems respectively.
More precisely their respective long time limits are : (1)
spatially homogeneous state, (2) fixed (steady) or periodic
structure and (3) chaotic pattern throughout space. The fourth
class of $\bf CA$ behaves in a much more complicated  manner and
was conjectured by Wolfram as capable for performing  universal
computation.
\newline
\indent Due to the complexity of cellular automata, numerical
computation of time evolution becomes the main scheme in $\bf CA$
studies. In \cite{Wol} some statistical approaches were introduced
to characterize quantitatively the patterns generated by $\bf CA$
evolution. Statistical quantities such as entropies and dimensions
can only provide average properties of cellular automata. However,
exact results are always needed in more elaborate discussions . To
make this point more explicit let us consider the effects of noise
on $\bf CA$. Due to the fact that all physical systems are coupled
to noisy environment, it is interesting to know if the above
classification of the attractors of $\bf CA$ remains intact under
the\newpage\noindent influence of noise. But this question can
only be answered by knowing the deterministic $\bf CA$ exactly,
and as a result the exact evaluation of the attractor without
noise is called for. Unfortunately such exact calculations only
exists for a few cases (A good review can be found in \cite{PhD}).
For example, the GKL automaton was proposed in the 80' \cite{Gac}
and surprisingly the detail analytic proof was completed about a
decade later \cite{Sa}. For finite $\bf CA$ with majority rules,
exact results have been obtained on finite lattice\cite{Gol}. It
is shown in \cite{Gol} that the attractors are either steady
states or periodic states of period two. It is important to note
that the result of \cite{Gol} contains the majority rules as a
special case. However, for a system of finite lattice, the
configuration space is finite and can only contain periodic final
states. Moreover, for infinite lattice, it is also known that the
attractors are steady states if the initial state is
finite\cite{Tch}(The finite initial condition will be defined in
the next section.). Since the above results are only true for
special configurations, the analysis for the attractors of
infinite $\bf CA$ with arbitrary initial configuration is still
lacking. It is the purpose of this work to fill this gap.
\newline
\indent
         In this work we provide the analytic results of infinite
one-dimensional cellular automaton. We investigate a subclass of
infinite $\bf CA$ and prove analytically that within this subclass
the chaotic and complex structures are excluded. As a result, the
only allowed attractors are homogenous, steady and periodic
states. The main tool for proving the existence of steady states
is the construction of a symbolic product among the fundamental
blocks which will be introduced later. In section 2, we briefly
introduce notations and formulation of $\bf CA$. In section 3 the
product rules are given and several lemmas are proved to set the
stage for the proof of the main result. The allowed attractors are
shown explicitly for the case of $k=2$ in section 4. Finally, a
brief discussion is given in section 5.\

\section{ CA review and notations}       % Enter section title between curly braces
\indent \ \ \ \ A cellular automaton is a spatial lattice of $N$
sites where $N$ can be finite or infinite. The state of each site
can be in one of $g$ states at time $t$. Each site follows the
same prescribed rules for up dating. For the rest of this work we
will only concentrate on one-dimensional $\bf CA$. The number of
neighborhood of each site is denoted by \boldmath $2k$ (In what
follows $k$ is always reserved specially for this context). The
$\bf CA$ starts out with arbitrary initial configuration which is
represented by $\bf
S(0)$=$\{s_1,s_2,s_3,\cdots,s_{\scriptscriptstyle N}\}$ where
$s_i$ can be in any one of the $g$ states. The configuration of
system at time $t$ is denoted by $\bf
S(t)$=$\{s_1(t),s_2(t),s_3(t),\cdots,s_{\scriptscriptstyle
N}(t)\}$. In this paper the state of each site is restricted to
$g=2$  such that $s_i\in\{1,-1\}$ and the updating rule is given
by the \textit{majority rule} defined as follows. The neighbors of
$s_i$ is defined as $\{s_{i+\alpha}|\alpha=\pm 1,\pm 2,\pm
3,\cdots,\pm k\}$. By introducing
\begin{equation}
 q_i=\sum_{{\scriptstyle \alpha=-k}}^{k}s_{i+\alpha},\end{equation} the updating rule can be expressed
as:\begin{equation}s_{i}(t+1)=\left\{\begin{array}{ccccc}1&&if
&{q_i}& {>0}\\-1&&&&{<0}\\s_{i}(t)&&&&{=
0}\end{array}\right.\end{equation} It is noted that the $k=1$ case
is also known as nearest neighbor {\bf CA}. In this work, $k$ is
arbitrary and hence our results go beyond nearest neighbor {\bf
CA}.
\newline\indent For finite one-dimensional $\bf CA$, the
lattice is arranged on a circle with periodic boundary conditions.
Such cellular automata have a finite number of states, and as a
result, after sufficient long time evolution the system must enter
the state which can either be homogenous, steady or periodic
state. Therefore the class 3 and 4 attractors do not exist in
finite $\bf CA$. In this paper exact results of a class of $\bf
CA$ are established for \textit{infinite extend}.

\section{Lemmas and product rules}       % Enter section title between curly braces
\indent \ \ \ \
  To represent the state of each site more conveniently, from
 now on, the binary state $\{1,-1\}$ will be graphically presented
as $\{+,-\}$ respectively. The one-dimensional $\bf CA$ starts out
with some initial configuration which is represented by $\bf
S(0)$=$\{s_1,s_2,s_3,\cdots,s_{\scriptscriptstyle N}\}$ where each
$s_i$ can take either $+$ or $-$. For any block of size $m$ (with
$m\geq k+1$) which contains only $+$($-$) site states is denoted
as ${A_m}({B_m})$ (In the following discussion, the subscript of a
block represents the block size. $m$ is reserved to denote block
size with $m\geq k+1$.):
\[\boldmath {A_m}=\{\underbrace{\cdots+++\cdots}_m\}\]
\[\boldmath{B_m}=\{\underbrace{\cdots---\cdots}_m\}.\]

\noindent Given $k$, any block of $l$ sites without containing
more than $k$ consecutive $+$ or $-$ site states is denoted by
\boldmath$X_l$. By using \boldmath$A_m$, \boldmath$B_m$ and
\boldmath$X_l$, any state can be symbolically decomposed as a
product of \boldmath$A_m$, \boldmath$B_m$ and \boldmath$X_l$, such
as {\boldmath$\{\cdots A_{m_1}X_{l_2}A_{m_3}B_{m_4}\cdots
X_{l_5}B_{m_6}\cdots\}$}. For instance, for $k=2$, the following
configuration can be expressed as:
\begin{eqnarray}
\lefteqn{\{----+-+--+-+++-+\}}\nonumber\\
&=&\{----\}\{+-+--+-\}\{+++\}\{-+\}\nonumber\\
&=&{\boldmath\{B_4X_7A_3X_2\}}\nonumber\end{eqnarray} Obviously,
this decomposition procedure is not unique. However, it is
important to note that this decomposition procedure is just a tool
for analyzing the state configuration at each time step and has
nothing to do with the dynamics of system. During evolution, the
size of each block may change and new block configurations will be
generate. Our main idea of proving the existence of attractors is
that \textit{the evolution of $CA$ can be reduced to the evolution
of these block variables}. For the subclass of cellular automata
considered in this paper, a fundamental properties which
constitutes the foundation of the final proof are given as lemmas.
\newtheorem{lemma}{Lemma}
\begin{lemma}
For $2k$ neighbors, the blocks {\boldmath$A_{k+1}$} and {\boldmath$B_{k+1}$} do not shrink during time evolution.
\end{lemma}
The proof of this lemma can easily be seen by noting that the
$q_i$ of each site in $A_{k+1}$($B_{k+1}$) is positive(negative)
or equal to zero. Hence these blocks can not shrink as the system
evolves. In fact the size of any \boldmath$A_{m}$ and
\boldmath$B_{m}$ may grow during time evolution. This is due to
the fact that all \boldmath$X_l$ configurations do not contain
more than $k$ consecutive $+$ or $-$ site states, therefore the
state of boundary site of \boldmath$X_l$ that is next to
\boldmath$A_m(B_{m})$ will switch sign as the system evolves. In
either case, the size of \boldmath$A_m(B_{m})$ grows. By assigning
a notion of parity to \boldmath$A_{m}$ as positive and
\boldmath$B_{m}$ as negative, this growing process expands in both
directions until both boundary sites meet another different parity
block, then the size of the block remains the same in subsequent
evolution. As a consequence, if the initial configuration contains
either \boldmath$A_{m}$ or \boldmath$B_{m}$, then the structure of
the attractor of these $\bf CA$ depends on the evolution of
\boldmath$X_l$. From the following lemma, any finite
\boldmath$X_l$ can only lead to homogeneous or steady states,
hence the chaotic state can never arise:
\begin{lemma}
Any finite \boldmath$X_l$ with either \boldmath$A_m$ or
\boldmath$B_m$ as boundaries on both sides with $m\geq k+1$ will
be eliminated during evolution.
\end{lemma}
Since the lemma is quite obvious we omit its proof. In fact this
lemma has been proved for any finite initial condition\cite{Tch}
which means that both \boldmath$A_m$ and \boldmath$B_m$ are of
infinite size. From \textbf{Lemma} $2$ one can see that these
requirements are not necessary. Therefore, for any initial state
without any block of \boldmath$X_\infty$, the attractors of $\bf
CA$ consist of the following configurations: homogeneous and
steady states. For example with any $k$, the state
\newpage\noindent
\[ S=\{B_{m_{1}}A_{m_{2}}B_{m_{3}}A_{m_{4}}\},\]
is a steady state which does not change during time evolution.
\newline \indent From above discussions, it is necessary to see if
the chaotic state can arise from any initial state containing
\boldmath$X_\infty$. By introducing and analyzing the combination
of products of the fundamental blocks which will be defined below,
the problem of \boldmath$X_\infty$ can be reduced to a problem of
finite \boldmath$X_l$.
\newline
\indent For any \boldmath$X_l$ with $l\geq 3(k+1)$, one can always
pick a smaller section \boldmath$X_{3(k+1)}$ ($\subset$
\boldmath$X_l$) and consider the evolution of this block. This
smaller section can then be split into three subblocks each of
which has a size of $k+1$: {\boldmath\[X_{3(k+1)}=
X_{k+1}Y_{k+1}Z_{k+1},\]} For example, when $k=2$, a configuration
{\boldmath$X_{3(k+1)}= X_{9}=\{--+-+-++-\}$} can be decomposed as:
{\boldmath\[X_{9}=\{--+-+-++-\}=\{--+\}\{-+-\}\{++-\}.\]}In this
way, any \boldmath$X_{3(k+1)}$ block can be built up by all the
possible \boldmath$X_{k+1}$'s which are defined as fundamental
blocks and we also denote them as $X$ without the subscript.
Obviously, by definition, the fundamental blocks do not contain
more than $k$ consecutive $+$ or $-$ site states. The number of
fundamental blocks is $2(2^k-1)$ and the total number of product
combinations of fundamental blocks for constructing
\boldmath$X_{3(k+1)}$ is bounded by $8(2^k-1)^3$ which can be
classified into five types: $\{XXX\}$, $\{YXX\}$, $\{XXY\}$,
$\{XYX\}$ and $\{XYZ\}$. Here, $X$, $Y$, $Z$ represent different
fundamental blocks respectively. In order to proof the existence
of the proposed attractors, one has to address how these possible
configurations evolve in time. It is obvious that the number of
product combinations grow rapidly as $k$ increases. However, by
realizing some relations among state configurations, the number of
product combinations needed to be considered can be reduced
effectively.
\newline
\indent It is easy to check that the product combinations of
$\{XXY\}$, $\{YXX\}$ and $\{XXX\}$ will produce $A_{m}$ or $B_{m}$
which will not shrink in subsequent evolution. For example, for
$k=2$, the product of two fundamental blocks $\{-+-\}\{-+-\}$ will
evolve to a configuration containing a block of \boldmath$B_{4}$
in the next time step. From {\bf Lemma} $1$ and $2$, one can
conclude that $\{-+-\}\{-+-\}$ can only contribute to the
transient states. This situation will be addressed in detail for
the case of $k=2$ in the next section. Therefore the cases
required further investigation are the product combinations of
$\{XYX\}$ and $\{XYZ\}$ types.
\newline
\indent As mentioned before, there exist relations among state
configuration in one-dimensional {\bf CA} such that the evolution
of the product combinations can be simplified. The simplification
is achieved by using two types of operations which were pointed
out in \cite{Mit}. For any configuration $\bf S(t)$=$\{\cdots
s_{n-1},s_n,\newline\noindent s_{n+1},\cdots\}$, the conjugate
state of $\bf S(t)$ is denoted as $\widetilde{\bf S}(t)$ defined
by flipping all the state of $s_i$. For example, the state $\bf
S(0)=\{-++-+-\}$ has a corresponding conjugate state
$\widetilde{\bf S}(0)=\{+--+-+\}$. Moreover, the state $\bf S(t)$
also has an image state associate to it. The image state $\bar{\bf
S}(t)$ of $\bf S(t)$ is defined by reversing sequence of the
states of sites of $\bf S(t)$. That is to say, for any $\bf S(t)$
with $N$ sites, $S(t)=\{s_1,s_2,\cdots,s_{{\scriptscriptstyle
N}-2},s_{{\scriptscriptstyle N}-1},s_{\scriptscriptstyle N}\}$,
the associate image state is
\begin{eqnarray*}
\bar{\bf
S}(t)&=&\{\bar{s}_1,\bar{s}_2,\cdots,\bar{s}_{{\scriptscriptstyle
N}-2}, \bar{s}_{{\scriptscriptstyle
N}-1},\bar{s}_{\scriptscriptstyle N}\}\\&=&\{s_{\scriptscriptstyle
N},s_{{\scriptscriptstyle N}-1},s_{{\scriptscriptstyle
N}-2},\cdots,s_2,s_1\}.
\end{eqnarray*}
For the $\bf S(0)$ considered above, the corresponding $\bar{\bf
S}(0)$ is $\{-+-++-\}$. It is important to realize that these
operations are preserved under time evolution. These results are
stated as theorems:
\newtheorem{theorem}{Theorem}
\begin{theorem}
If $\widetilde{\bf S}(t)$ is the conjugate state of $\bf S(t)$,
then $\widetilde{\bf S}(t+1)$ is the conjugate state of $\bf
S(t+1)$.
\end{theorem}
Proof: Under time evolution
\[s_i(t+1)=sgn\{\sum_{\alpha=1}^{k}(s_{i+\alpha}+s_{i-\alpha})\},\]
where \textit{\textbf{sgn}} is the conventional sign-function and
if $\sum(s_{i+\alpha}+s_{i-\alpha})=0$, then $s_i(t+1)=s_i(t)$.
The evolution of $\widetilde{\bf S}(t)$ is given by:
\begin{eqnarray*}
\tilde{s}_i(t+1)&=&sgn\{\sum_{\alpha=1}^{k}[\tilde{s}_{i+\alpha}(t)+\tilde{s}_{i-\alpha}(t)]\}\\
             &=&sgn\{\sum_{\alpha=1}^{k}[(-1)s_{i+\alpha}(t)+(-1)s_{i-\alpha}(t)]\}\\
             \\&=&sgn\{(-1)\sum_{\alpha=1}^{k}[s_{i+\alpha}(t)+s_{i-\alpha}(t)]\}\\
             &=&(-1)s_i(t+1).
\end{eqnarray*}
When $\sum(s_{i+\alpha}+s_{i-\alpha})=0$, $s_i(t+1)=s_i(t)$, then
\[\tilde{s}_i(t+1)=(-1)s_i(t)=\tilde{s}_i(t).\]
That is to say, if $s_i(t)$ remains unchanged, so does
$\tilde{s}_i(t)$. The proof is complete.
\begin{theorem}
If $\bar{\bf S}(t)$ is the image state of $\bf S(t)$, then $\bar{\bf S}(t+1)$ is the image state of $\bf S(t+1)$.
\end{theorem}
Proof: The proof is similar to the {\bf theorem} $1$. The image of
$S(t)$ is given by
\[\bar{S}(t)=\{\bar{s}_1,\bar{s}_2,\cdots,\bar{s}_{{\scriptscriptstyle
N}-2},\bar{s}_{{\scriptscriptstyle
N}-1},\bar{s}_{\scriptscriptstyle N}\}\] with
$\bar{s}_i=s_{{\scriptscriptstyle N}-i}$. At the next time step,
one has
\begin{eqnarray*}
\bar{s}_i(t+1)&=&sgn\{\sum_{\alpha=1}^{k}[\bar{s}_{i+\alpha}(t)+\bar{s}_{i-\alpha}(t)]\}\\
             &=&sgn\{\sum_{\alpha=1}^{k}[s_{{\scriptscriptstyle
N}-(i+\alpha)}(t)+s_{{\scriptscriptstyle
N}-(i-\alpha)}(t)]\}\\
             &=&sgn\{\sum_{\alpha=1}^{k}[s_{({\scriptscriptstyle
N}-i)+\alpha}(t)+s_{({\scriptscriptstyle
N}-i)-\alpha}(t)]\}\\
             &=&s_{{\scriptscriptstyle
N}-i}(t+1).
\end{eqnarray*}
It is also true that if $s_{{\scriptscriptstyle
N}-i}(t+1)=s_{{\scriptscriptstyle N}-i}(t)$, then
$\bar{s}_i(t+1)=\bar{s}_i(t)$. Thus the image operation is
persevered under time evolution.
\newline
\indent The method of establishing the existence of steady states
is to show that all \boldmath$X_{3(k+1)}$ can produce $A_m$ or
$B_m$ as the system evolves. Once $A_m$ or $B_m$ occurred, by
using {\bf lemma} $1$ and $2$, the steady state is obtained. From
now on $A_m$ and $B_m$ will be referred as the \textit{uniform
blocks}. However, one of the important properties of the
conjugation operation is that the blocks $A_{m_a}$ and $B_{m_b}$
with $m_a=m_b$ form a conjugate pair. This fact is very useful in
simplifying the analysis of the attractors.
\newline
\indent It is important to realize that the conjugation procedure
divides the set of all possible product combinations of
\boldmath$X_{3(k+1)}$ into two mutually conjugate classes. Any
\boldmath$X_{3(k+1)}$ will belong to one of these two classes
unless it is self-conjugate. The mutual conjugation between
$A_{m_a}$ and $B_{m_a}$ reduces the analysis on just one class.
This reduction can be seen from the fact that if $Q_{3(k+1)}$
which is in either of classes produces a block $A_{m_a}$ during
evolution, then by applying {\bf theorem} $1$, the conjugate block
$\tilde{Q}_{3(k+1)}$ which belongs to the other conjugate-class
will produce $B_{m_a}$. As a\newpage\noindent result, only one of
the classes denoted by $\Gamma$ requires further discussion. (The
other conjugate-class is denoted by $\tilde{\Gamma}$.) On
$\Gamma$, further reduction can also be obtained by considering
the imaging operation. By considering two product combinations in
$\Gamma$ which form the image pair, if one of the pair produces
either of the uniform blocks during evolution, then the other one
results in a corresponding block respectively. Hence, one of these
two product combinations can be eliminated without losing
generality. However, in $\Gamma$, there are some product
combinations that the image of them appear in $\tilde{\Gamma}$.
Then the conjugation of these image product combinations are also
in $\Gamma$ and can also be eliminated. This fact can easily be
seen by using the combined results of {\bf theorems 1} and $2$.
The subset of $\Gamma$ obtained after the above reductions is the
set of product combinations which require detail analysis of the
product rules. In the next section a detail discussion on the
product rules for $k=2$ will be presented.
\section{The $k=2$ Model}       % Enter section title between curly braces
\indent \ \ \ For $k=2$, there are six fundamental blocks:
\begin{equation}
{\boldmath
\begin{array}{cccc}

 E=\{+-+\}&G=\{++-\}&H=\{-++\}\\
 F=\{-+-\}&I=\{--+\}&J=\{+--\},

\end{array}
}\end{equation} From (3), it is obvious that
\begin{equation}
{\boldmath\{E,G,H\}}
\end{equation}
are the conjugate-blocks of $\{F,I,J\}$ respectively. Moreover,
the image operation produces the image pairs, ($G,H$) and ($I,J$).
All $X_9$ can be built up by the product combinations of (3) and
altogether one has at most $216$ product combinations. In
considering the evolution of any $X_{9}$, as mentioned before, it
can be checked explicitly that the product combinations of
$\{XXX\}$, $\{XXY\}$ and $\{YXX\}$ types can produce the uniform
block in the next time step and lead to the steady state. Note
that this fact is also true for arbitrary $k$. Therefore, the main
goal of the following discussion is to prove the existence
attractors generated from $\{XYX\}$ and $\{XYZ\}$.
\newline
\indent Let us start from the reduction procedures. It is noted
that there are some products where $A_m$ or $B_m$ arises
automatically as two fundamental blocks are grouped together.
These products are excluded from the initial state with
configuration $X_{\infty}$. For example, the product of $E$ and
$G$ is {\boldmath
\[\{EG\}=\{+-+++-\}=\{X_2A_3X_1\}.\]}
The complete list of these particular results is as follows:
\begin{equation}
{\boldmath \{EG,FI,GI,IG,HG,JI,HJ,JH,HE,JF\}. }\end{equation}
Moreover, there are also four products which can evolve to uniform
blocks within one step and without referring to their neighboring
fundamental block. These products are $\{EH,GE,FJ,IF\}$, and will
also be omitted in the following discussion. As a result, the
remaining products of two fundamental blocks are:
\begin{equation}{\boldmath\{EF,EI,EJ,GF,GH,GJ,HF,HI,FE,FG,FH,IE,IJ,IH,JE,JG\}.}
\end{equation}
This set can be divided into two subsets by conjugation relation.
They are symbolically presented as follows for expressing the
product relation more effectively:
\begin{equation}
{\boldmath
E \mapsto \left\{\begin{array}{c}
F\\I\\J
\end{array}\right\}
, G \mapsto \left\{\begin{array}{c}
F\\H\\J
\end{array}\right\}
, H \mapsto \left\{\begin{array}{c}
F\\I
\end{array}\right\}
}\end{equation}\newpage\noindent and
\begin{equation}
{\boldmath
F \mapsto \left\{\begin{array}{c}
E\\G\\H
\end{array}\right\}
, I \mapsto \left\{\begin{array}{c}
E\\J\\H
\end{array}\right\}
, J \mapsto \left\{\begin{array}{c}
E\\G
\end{array}\right\}.
}\end{equation} Therefore, if the product combinations generated
from (7) are in the class $\Gamma$, then the ones generated from
(8) must belong to the corresponding conjugate-class
$\tilde{\Gamma}$. Keeping these products and the conjugate
relation between (7) and (8) in mind, the structure of attractors
can be shown explicitly.
\newline \indent Starting from $\{XYX\}$, because of the conjugate
relation between (7) and (8), one only need to use the group (7)
to construct the product combinations of such type. By further
considering the imaging operations, the product combinations can
be reduced to only $4$ cases:
\begin{equation} {\boldmath\{EFE\},\{EIE\},\{GFG\},\{GJG\}}.
\end{equation}
These combinations by itself do not contain enough information to
justify that the system will produce uniform blocks at the next
time step. However, it will be shown latter that these product
combinations can lead to either the steady or periodic states.
\newline\indent
For the $\{XYZ\}$ type, the analysis can be proceeded by
concentrating on the ones in $\Gamma$. There are $14$ product
combinations required further consideration. By using the imaging
operation for further reduction, the number of remaining product
combinations is $10$. Moreover, by combining the procedures of
conjugation and imaging together, only $8$ combinations remain:
\begin{equation}
{\boldmath
 \{EFG\},\{EFH\},\{EIJ\},\{EIH\},\{EJG\},\{GFH\},\{GHI\},\{HFG\}.}
\end{equation}
It is easy to check that except for the case of $\{GHI\}$, the
rest of (10) will generate uniform block within one or two steps.
For $\{GHI\}$, in order to check whether it can generate uniform
blocks, it is necessary to take into account the effects of its
neighboring blocks as the system evolves. Since ${\{EG, IG, HG\}}$
is a subset of (5) which already contain the uniform blocks, the
neighboring blocks of the leftmost side of $\{GHI\}$ should not
include any of $\{E,I,H\}$. Similarly, for the rightmost side of
$\{GHI\}$, one should not include any of $\{F, G, J\}$. Thus, the
following combinations require separate investigation:
\begin{equation}
 \{FGHIE\},\{FGHIH\},\{FGHIJ\},\{JGHIE\},\{JGHIH\},\{JGHIJ\}.
\end{equation}
\noindent By using the fact that the conjugate of $\{FGH\}$ is
$\{EIJ\}$ which is already in (10), that is to say any block in
(11) which contains $\{FGH\}$ will generate the uniform block.
This discussion also apply to $\{JGHIE\}$ and $\{JGHIH\}$.
However, the remaining case $\{JGHIJ\}$ does not produce the
uniform block. This particular block suggests a periodic
configuration. In fact the infinite sequence of
$\boldmath\{JGHI\}$ is a periodic state:
\[\begin{array}{ccccc}
\{\cdots JGHIJGHI \cdots\}\\
\downarrow \\
\boldmath\{\cdots HIJGHIJG \cdots\}\\
\downarrow \\
\boldmath\{\cdots JGHIJGHI \cdots\}
\end{array}\]
Furthermore, by checking explicitly, any other neighboring block
attached to $\{JGHIJ\}$ always create a uniform block within two
steps.
\newline \indent
Similar analysis can now be applied for analyzing the product
combinations in (9). This can be illustrated by the case of
$\{EFE\}$. According to the rules given by (7) and (8), the only
neighboring blocks of $\{EFE\}$ which do not generate uniform
blocks is the infinite sequence: $\{\cdots FEFEFEFE\cdots\}$. This
configuration is a steady state. By taking this block
configuration with infinite sequence, the remaining product
combinations in (9) are periodic states. For instance, considering
$\{EIE\}$, the infinite sequence is $\{\cdots EIEIEI \cdots\}$.
Under time evolution, this state oscillates:\\ {\boldmath
\[\begin{array}{ccccc}
\{\cdots EIEIEI \cdots\}\\
\downarrow \\
\boldmath\{\cdots GFGFGF \cdots\}\\
\downarrow \\
\boldmath\{\cdots EIEIEI \cdots\}
\end{array}\]
}The other periodic state is: \boldmath
\[ \{\cdots GJGJGJ \cdots\} \Longleftrightarrow \{\cdots IHIHIH \cdots\} \]
On the other hand, any other neighboring block attached to any
blocks in (9) will produce uniform blocks by checking directly. In
passing we note that the periodic attractors are period two as the
finite $\bf CA$.
\newline\indent To summarize, for $k=2$, from above analysis, we reduce the
number of product combinations from $216$ down to $12$ ((9)+(10))
and it is shown that all attractors of the $k=2$ model are
consisted of three different types:
\begin{enumerate}
\item The spatially homogeneous states such as $A_{\infty}$ and
$B_{\infty}$.
\item The steady states of the form ${\boldmath\{... EFEF ...\}}$ and
${\boldmath\{... A_{m_{i}}B_{m_j}A_{m_{p}}B_{m_{q}} ...\}}$ with
$\forall m_a\geq k+1$.
\item The periodic states which includes
${\boldmath\{...JGHIJGHI...\}}$, ${\boldmath\{...EIEI...\}}$ and
${\boldmath\{...GJGJ...\}}$.
\end{enumerate}
Therefore, the chaotic state and complex structure state do not
exist in this cellular automaton.

\section{Discussions and conclusions}       % Enter section title between curly braces
\indent \ \ \ \ By realizing fundamental blocks structure of the
{\bf CA} model with majority rule, it is shown in this paper that
this model has certain nice properties which can be used to
establish the structure of the attractors. It is shown explicitly
that, with $k=2$, for arbitrary initial configuration the only
allowed attractors are the spatially homogeneous, steady and
periodic states. Thus, the chaotic state and complex structure
state referred in Wolfram's classification \cite{Wol} are
excluded. Our method also provides explicit enumeration of all
periodic states which are in fact period two as in the case of
finite lattice \cite{Gol}.
\newline
\indent Although the proof was established for $k=2$, this
approach is general enough for showing the result of arbitrary
$k$. The approach is based on the fact that the uniform blocks
$A_{k+1}$ and $B_{k+1}$ grow during evolution. Furthermore, by
using the fact that any initial condition can be expressed as
block products, therefore the problem of deriving the attractors
turns into the problem of showing that any \boldmath$X_{3(k+1)}$
can generate the uniform blocks. Even though the number of product
combinations of $X_{3(k+1)}$ increases with $k$, by using the
conjugation and imaging procedures, the amount of explicit
checking on the product analysis is greatly reduced.  This
approach has also been done for the case of $k=3$ but the analysis
is too lengthy to be presented here and hence omitted. Thus, this
work provides a complete analysis of the attractors of infinite
{\bf CA} with majority rules. Our results hold for any initial
configuration in contrast to the results of previous works which
assume either finite initial states or finite lattice. One can
conclude that for this subclass of {\bf CA} model, the only
allowed attractors are homogenous, steady and periodic states.
\newline
\newline \noindent{\bf Acknowledgement.} This work was supported
by {\bf NSC90-2112-M-006-014}.

% Set the ending of a LaTeX document
\end{document}